\pdfoutput=1 

\documentclass[final,1p,times]{elsarticle}

\usepackage{epsfig}

\usepackage{amssymb}

\usepackage{algorithmic} 
\usepackage{stackrel} 

\journal{Communications in Computational Physics}

\begin{document}

\begin{frontmatter}

\title{An Algorithm for the Stochastic Simulation of Gene Expression  
and Heterogeneous Population Dynamics}

\author{Daniel A. Charlebois$^{a,b,1}$, Jukka Intosalmi$^{c,d}$, Dawn  
Fraser$^{a,b}$, Mads K$\ae$rn$^{a,b,e,1}$}

\address[label1]{Department of Physics, University of Ottawa, 150  
Louis Pasteur, Ottawa, Ontario, K1N 6N5, Canada.}
\address[label2]{Ottawa Institute of Systems Biology, University of  
Ottawa, 451 Symth Road, Ottawa, Ontario, K1H 8M5, Canada.}
\address[label3]{Department of Mathematics, Tampere University of  
Technology, P.O. Box 553, 33101 Tampere, Finland.}
\address[label4]{Department of Signal Processing, Tampere University  
of Technology, P.O. Box 553, 33101 Tampere, Finland.}
\address[label5]{Department of Cellular and Molecular Medicine,  
University of Ottawa, 451 Symth Road, Ottawa, Ontario, K1H 8M5, Canada.}

\begin{abstract}
We present an algorithm for the stochastic simulation of gene  
expression and heterogeneous population dynamics. The algorithm  
combines an exact method to simulate molecular-level fluctuations in  
single cells and a constant-number Monte Carlo method to simulate time-dependent
statistical characteristics of growing cell populations. To  
benchmark performance, we compare simulation results with steady-state  
and time-dependent analytical solutions for several scenarios,  
including steady-state and time-dependent gene expression, and the  
effects on population heterogeneity of cell growth, division, and DNA  
replication. This comparison demonstrates that the algorithm provides  
an efficient and accurate approach to simulate how complex biological  
features influence gene expression. We also use the algorithm to model  
gene expression dynamics within `bet-hedging' cell populations during  
their adaption to environmental stress. These simulations indicate  
that the algorithm provides a framework suitable for simulating and  
analyzing realistic models of heterogeneous population dynamics  
combining molecular-level stochastic reaction kinetics, relevant  
physiological details and phenotypic variability.
\end{abstract}


\footnotetext[1]{
Corresponding authors. Tel.: +1 613 562 5800 (Ext. 8691); Fax: (+1) 613  
562 5636.\\*
\textit{E-mail addresses:} daniel.charlebois@uottawa.ca (Daniel Charlebois); mkaern@uottawa.ca (Mads K$\ae$rn).}

\end{frontmatter}

\section{Introduction}

Stochastic mechanisms play key roles in biological systems since the  
underlying biochemical reactions are subject to molecular-level   
fluctuations (see e.g. \cite{Kaern,Arkin}). Chemical reactions are  
discrete events occurring between randomly moving molecules.  
Consequently, the timing of individual reactions is nondeterministic  
and the evolution of the number of molecules is inherently noisy. One  
example of particular importance is the stochastic expression of gene  
products (mRNA and protein)  
\cite{Kaern,Kaufmann,Maheshri,Paulsson,Arkin}. Here, molecular-level  
fluctuations may cause genetically identical cells in the same  
environment to display significant variation in phenotypes, loosely  
defined as any observable biochemical or physical attribute. While  
such noise is generally viewed as detrimental due to reduced precision  
of signal transduction and coordination, several scenarios exist where  
noise in gene expression may provide a fitness advantage (see Fraser  
and K{\ae}rn \cite{Fraser} for a review). For example, it has been  
proposed that a cell population may enhance its ability to reproduce  
(fitness) by allowing stochastic transitions between phenotypes to  
increase the likelihood that some cells are better positioned to  
endure unexpected environmental fluctuations \cite{Oudenaarden}.

Due to the importance of noise in many biological systems, models  
involving stochastic formulations of chemical kinetics are  
increasingly being used to simulate and analyze cellular control  
systems \cite{Gillespie3}. In many cases, obtaining analytical  
solutions for these models are not feasible due to the intractability  
of the corresponding system of nonlinear equations. Thus, a Monte  
Carlo (MC) simulation procedure for numerically calculating the time  
evolution of a spatially homogeneous mixture of molecules is commonly  
employed \cite{Gillespie1,Gillespie2}. Among these procedures, the  
Gillespie stochastic simulation algorithm (SSA) is the \textit{de-facto}
standard for simulating biochemical systems in situations where  
a deterministic formulation may be inadequate \cite{Gillespie1}. The  
SSA tracks the molecular number of each species in the system as  
opposed to the variation in concentrations in the deterministic  
framework. Hence, high network complexity, large separation of time-scales
and high molecule numbers can result in computationally  
intensive executions. Another challenge is the need for simulating  
cell populations. In many cases, gene expression is measured for  
10-100 thousand individuals sampled from an exponentially growing  
culture of continuously dividing cells. While the dynamics of  
individual cells can be appropriately simulated by disregarding  
daughter cells, repeating such simulations for a fixed number of cells  
may not capture population variability arising from asymmetric  
division, for example, or age-dependent effects. The alternative,  
tracking and simulating all cells within the population, is  
intractable beyond a few divisions due to an exponential increase in  
CPU demands as a function of time \cite{Mantzaris2}.

Here, we present a flexible algorithm to enable simulations of  
heterogeneous cell population dynamics at single-cell resolution.  
Deterministic and Langevin approaches to account for changes in  
intracellular content and the constant-number MC method  
\cite{Lin,Smith} were previously been combined to simulate and analyze  
gene expression across cell populations \cite{Mantzaris1,Mantzaris2}.  
In these studies, extrinsic heterogeneity associated with stochastic  
division and partitioning mechanisms, and intrinsic heterogeneity  
associated with molecular reaction kinetics were considered. Our  
algorithm, which combines the exact SSA for single-cell molecular- 
level modeling and a constant-number MC method for population-level  
modeling, is designed to incorporate user-defined biologically  
relevant features, such as gene duplication and cell division, as well  
as single cell, lineage and population dynamics at specified sampling  
intervals. Additionally, the SSA, which can be replaced by approximate  
methods if desired, is implemented within a shared-memory CPU  
parallelization framework to reduce simulation run-times. The emphasis  
of our study is to validate the accuracy of the method by directly  
comparing simulated results to the analytical solutions of models  
describing increasingly realistic biological features. Our results  
indicate that combining the SSA and the constant-number MC provides an  
efficient and accurate approach to simulate heterogeneous population  
dynamics, and a reliable tool for the study of population-based models  
of gene expression incorporating physiological detail and phenotypic  
variability.

This paper is organized as follows: Sections 2 and 3 briefly introduce  
the SSA and the constant-number MC method, respectively. The developed  
algorithm is described in Section 4. Section 5 provides the results of  
the benchmarking against analytical results. Finally, in Section 6, we  
demonstrate the applicability of the algorithm to more complex  
contexts by demonstrating that it can quantitatively reproduce  
experimental measurements of gene expression dynamics within `bet- 
hedging' cell populations during their adaption to environmental  
stress. The work is summarized in Section 7.

\section{Stochastic Simulation Algorithm}
The physical basis of the stochastic formulation of chemical kinetics  
is a consequence of the fact that collisions in a system of molecules  
in thermal equilibrium is essentially a random process  
\cite{Gillespie2}. This stochasticity is correctly accounted for by  
the Gillespie SSA, a MC procedure to numerically simulate the time  
evolution of chemical and biochemical reaction systems. While based on  
an assumption of intracellular homogeneity and mass-action kinetics,  
it is the $\textit{de-facto}$ standard for simulations of gene  
expression. In the Direct Method Gillespie SSA, $M$ chemical reactions  
$R_{1},\ldots,R_{M}$ with rate constants $c_{1},...,c_{M}$ among $N$  
chemical species $X_{1},...,X_{N}$, are simulated one reaction event  
at a time. The next reaction to occur (index $\mu)$ and its timing ($ 
\tau$) are determined by calculating $M$ reaction propensities  
$a_{1},...,a_{M}$, given the current number of molecules of each of  
the $N$ chemical species, to obtain an appropriately weighted  
probability for each reaction. It can be implemented via the following  
pseudocode \cite{Gillespie1,Gillespie2}:\\*
\begin{algorithmic}[1]
        \IF{$t < t_{end}$ and $\alpha_{M} = \sum_{v=1}^M a_v \neq 0$}
                \FOR{$i=1,M$}
                        \STATE Calculate $a_{i}$ and $\alpha_{i} = \sum_{v=1}^i a_v$
                \ENDFOR
                \STATE Generate uniformly distributed random numbers ($r_1$,$r_2$)
                \STATE Determine when ($\tau=\ln(1/r_1)/\alpha_{M}$) and which ($ 
\min \{~\mu~|~\alpha_{\mu} \geq r_2\alpha_{M}\}$) reaction will occur
                \STATE Set $t=t+\tau$        
                \STATE Update $\left\{X_{i}\right\}$          
        \ENDIF
\end{algorithmic}
\mbox{}

The SSA can be augmented to incorporate biologically relevant  
features, such as changes in the volume of the cell during growth, the  
partitioning of cell volume and content at division and DNA  
replication (see e.g. \cite{Adalsteinsson,Lu,Ribeiro}). Changes in  
cell volume may have significant effects on reaction kinetics. First  
order reactions have deterministic rate constants ($w_{M}$) and  
stochastic rate constants ($c_{M}$) that are equal and independent of  
volume \cite{Kierzek}. However, for higher order reactions, it is  
necessary to incorporate cell volume $V(t)$ into the reaction  
propensities in order to perform an exact simulation. For example, the  
stochastic rate constant for a bimolecular second order reaction  
$R_{\mu}$ at time $t$ is given by
\begin{equation}\label{rates_eq1}
c_{\mu}=\frac{w_{\mu}}{N_{A}V_{k}(t)},
\end{equation}
where $N_{A}$ is Avogadro's number. Therefore, in the SSA, the rates  
of higher-order reactions must be scaled appropriately by the current  
cell volume before calculating propensities. This procedure has  
previously been demonstrated to provide a satisfying approximation as  
long as the kinetic time-scale is short compared with the cellular  
growth rate \cite{Lu}. Typically, the volume of each cell $k$ is  
modeled using an exponential growth law
\begin{equation}\label{growth_eq}
V_{k}(t_{div})=V_{0}\exp\left[\ln(2) \left(\frac{t_{div}}{\tau_{0}} 
\right)\right],
\end{equation}
where $V_{0}$ is the cell volume at the time of its birth, $t_{div}$  
is the time and $\tau_0$ is the interval between volume doublings.  
This functional form allows for the description of dilution as a first- 
order decay process within a deterministic model of intracellular  
concentrations.

Once the SSA incorporates a continuously increasing cell volume, it is  
necessary also to specify rules that govern cell division. One option  
is `sloppy cell-size control' \cite{Tyson} where the cell division is  
treated as a discrete random event that take place with a volume-dependent
probability. Another simpler option is to assume that  
division occurs once the cell has exceeded a critical size $V_{div}$  
corresponding to one doubling of its initial volume, $V_{div}  
=2V_{0}$. The volume doubling time $\tau_0$ then becomes cell division  
time and $t_{div}$ becomes the time since the last division. When cell  
division is triggered, i.e. when $V_{k}(t_{div}) \geq V_{div}$,  
additional rules must be specified to model the partitioning of  
cellular content between mother and daughter cells. For example,  
asymmetric cell division can be modeled by setting $V_{daughter} <  
V_{mother}$. The molecules of the cell can then be partitioned  
probabilistically between the two volumes  
\cite{Kierzek,Swain2,Swain3,Swain}.

The SSA can accommodate additional discrete events. For example, the  
G2/M cell cycle checkpoint which ensure proper duplication of the  
cell's chromosomes before division, can be modeled by defining a  
variable representing the completion of DNA replication such that cell  
division is delayed until the DNA content of the cell has doubled. The  
replication of individual genes, which doubles the maximum rate of  
gene transcription by doubling the number of corresponding DNA  
templates, can be modeled as a discrete event that occurs at a fixed  
time $t_{rep}$ after cell division, i.e. when $t_{div} \geq t_{rep}$,  
or as a random event that occurs with some variable probability. In  
both cases, the DNA-replication event can be placed in a cell-specific  
stack of future events that is compared against $t_{div}$ (or $t$ in  
the above pseudocode) following each SSA step. Events in the stack  
scheduled to occur before this time are then executed and removed from  
the stack. This can be incorporate into the above pseudocode by  
inserting the following two lines:\\*\\*
8a: \textbf{if} $length(t_{event})\ge 0$ \textbf{then} (there are  
scheduled events)\\*
8b: \textbf{if} $t>t_{event}(i)$ \textbf{then} execute event(i) and  
delete $t_{event}(i)$ from stack\\*\\*
This approach also provides a convenient basis for simulating the  
effects of time-delays \cite{Ribeiro,Roussel}.

We note that the exact SSA can be extremely computationally intensive  
since the step size $\tau$ becomes very small when the total number of  
molecules is high or the fastest reaction occurs on a time-scale that  
is much shorter than the time-scale of interest. It therefore useful  
to develop techniques that can be used to speed up the simulation.  
This can be done, for example, using approximate methods such as the  
tau-leaping procedure in which each time step $\tau$ advances the  
system through possibly many reaction events \cite{Gillespie4}.  
Additionally, since many independent runs are required to compute  
population statistics, parallel computing can be used to further  
optimize simulation run-times.

\section{Constant-Number Monte Carlo} \label{cnmc}

Implementations of the modified SSA that track only one of the two  
cells formed during cell division may introduce artifacts in the  
calculation of population characteristics in the presence of  
significant phenotypic variability among cells. For example, gene  
expression capacity and division time may depend on chronological age;  
old cells may express genes at a reduced rate, and daughter cells may  
need to mature before they can reproduce. In addition, reproductive  
rates may be influenced by the accumulation of genetic mutations  
within a specific cell lineage or by the current levels of gene  
expression within individual cells. To simulate stochastic models of  
gene expression incorporating such features, it is necessary to couple  
the SSA with simulation techniques used in studies of population  
dynamics.

The population balance equation (PBE) is a mathematical statement of  
continuity that accounts for all the processes that generate and  
remove particles from a system of interest \cite{Ramkrishna},  
including individual members of a population \cite{Smith}. In a  
general molecular-dynamics framework, the PBE contains terms due to  
nucleation, coagulation and fragmentation, and so forth, and is  
mathematically represented by an integro-differential equation that  
typically must be solved numerically to obtain particle size  
distribution and densities as a function of time \cite{Smith}. Due to  
the integro-differential nature of the problem, discretization of the  
size distribution is required. This is problematic because features of  
the distribution are not known ahead of time and may change during  
growth \cite{Kostoglou,Smith}. To resolve discretization problems that  
hinder the direct integration of the PBE, one can use MC methods to  
sample a finite subset of a system in order to infer its properties  
and study finite-size effects, spatial correlations, and local  
fluctuations not captured by a mean field approximation  
\cite{Gillespie4,Lin,Ramkrishna,Smith}. Furthermore, a MC method is  
appropriate as its discrete nature adapts itself naturally to growth  
processes involving discrete events, and can simulate growth over  
arbitrary long times with finite numbers of simulation particles while  
maintaining constant statistical accuracy~\cite{Lin}.

In order to construct a reliable and efficient algorithm to simulate  
biological cell populations, a constant-number MC method is adopted to simulate the birth-death processes that  
take place within such populations \cite{Lin,Mantzaris1,Mantzaris2,Smith}. This approach permits modeling  
of growing populations using a fixed number of cells while avoiding  
the alternative (i.e. an infinitely growing population) by sampling $N 
$ particles representing the population as a whole. It essentially  
amounts to contracting the physical volume represented by the  
simulation to continuously maintain a constant number of cells  
\cite{Lin}. The constant-number MC approach has been successfully applied
to a variety of non-biological particulate processes \cite{Lee,Lin,Smith}
as well as cell population dynamics \cite{Mantzaris1,Mantzaris2}.

In our implementation of the constant-number MC, we keep track of  
individual mother and daughter cells in two separate arrays. Each time  
a cell divides, the daughter cell is placed in the daughter array and  
the time of birth recorded. Then, at specified intervals, cells within  
the mother array are replaced one at a time, with the oldest daughter  
cells being inserted first. Because every mother cell is equally  
likely to be replaced during the sample update, the size distribution  
of the population remains intact for sufficiently large populations~ 
\cite{Smith}. In our case, the size distribution corresponds to the  
distribution of cell ages (or volumes) across the population.

The constant-number MC method can be represented by the following  
pseudocode:\\
\begin{algorithmic}[1]
        \IF{$t > t_{restore}$ and $NC_{daughter} \geq 1$}
                \FORALL{$NC_{daughter}$}
                        \STATE Randomly select mother cell
                        \STATE Replace mother cell with oldest available daughter cell
                \ENDFOR
        \ENDIF
\end{algorithmic}
Here, $t_{restore}$ is the interval between population updates and  
$NC_{daughter}$ the number of daughter cells born since the last  
update. To avoid simulating the daughters of daughter cells,  
$t_{restore}$ is chosen such that mother cells divide at most once,  
and daughter cells not at all, during a particular $t_{restore }$  
interval.

\section{Algorithm} \label{sec:algorithm}
\begin{figure}[h]
\begin{center}
\includegraphics*[width=8cm,height=10cm]{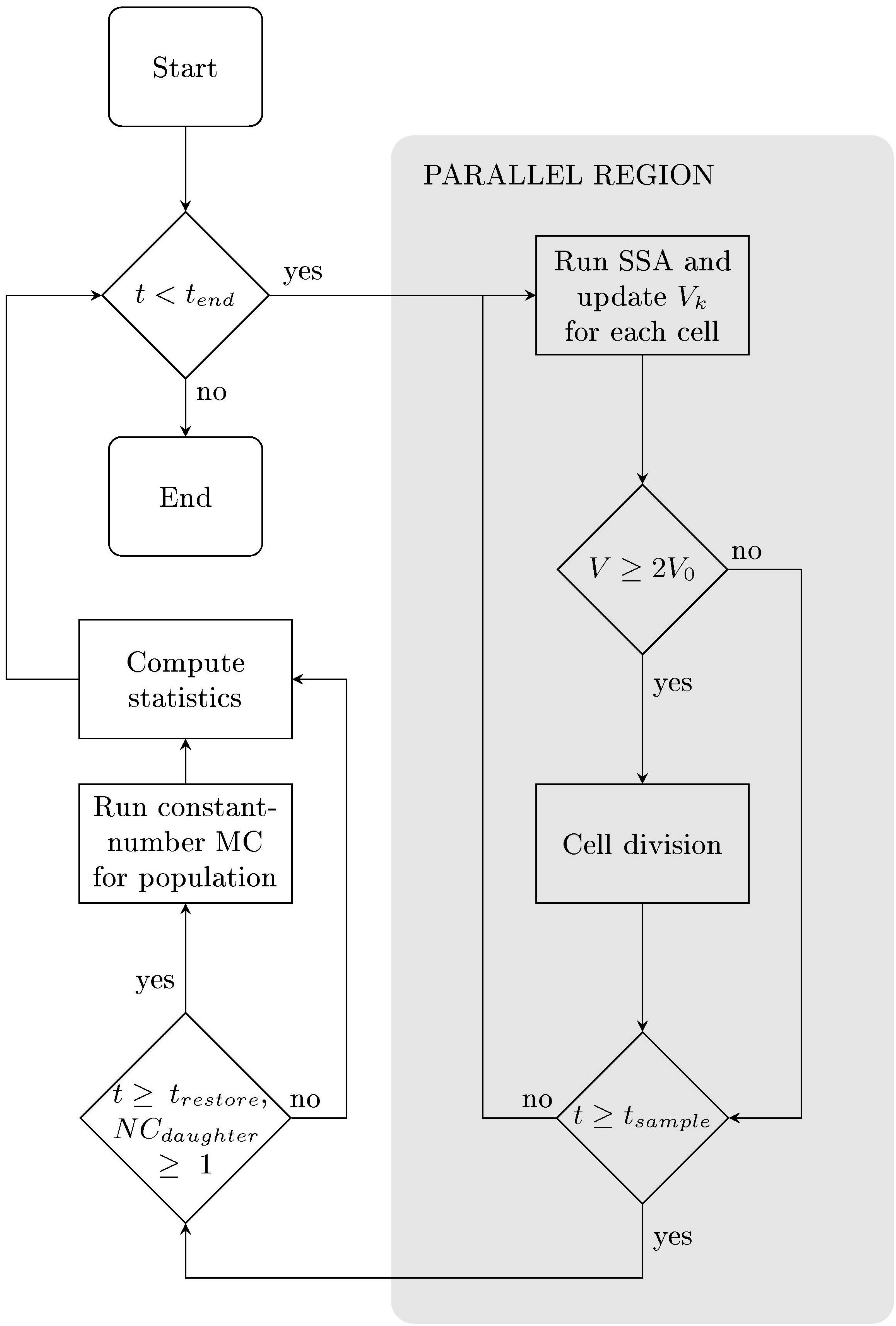}
\end{center}
\caption{\protect
     Flow diagram of the present algorithm for the parallel stochastic  
simulation of gene expression and heterogeneous population dynamics.
} \label{flowchart}
\end{figure}

Simulations are carried out using an initial population distribution,  
where gene expression in each cell is described by a user defined set  
of equations, and population statistics are obtained at a specified  
sampling interval. Here, stochastic simulation is carried out using  
the Gillespie direct method \cite{Gillespie1,Gillespie2}, however any  
stochastic simulation method can be implemented. Parallelism is  
implemented across the simulation (see Fig.~\ref{flowchart} and  
pseudocode in this section), as a large number of independent  
simulations need to be performed when simulating the dynamics of a  
cell population, in a shared memory multiprocessor environment.

The algorithm can be expressed by the flow diagram (Fig.~ 
\ref{flowchart}) and the following pseudocode: \\
\begin{algorithmic}[1]
\WHILE{$t < t_{end}$}
        \STATE \emph{begin parallel region}
        \FORALL{$NC_{population}$ such that $t < t_{sample}$}
                \STATE Gillespie SSA (see pseudocode in Section 2)
                \STATE Update $V_{k}$
                \STATE Execute events in stack with $t_{event} < t_{div}$
                \IF{$V_{k}(t_{div}) \geq V_{div}$}
                        \STATE Execute cell division
                        \STATE Increment $NC_{daughter}$
                \ENDIF        
        \ENDFOR
        \STATE Update $t_{sample}$
        \STATE \emph{end parallel region}  
        \STATE Execute constant-number MC (see pseudocode in Section 3)      
        \STATE Compute statistics
\ENDWHILE
\end{algorithmic}
Here, $NC_{population}$ is the total number of cells in the  
population, $V_{k}$ the volume of cell $k$, and $t_{sample}$ the user  
defined population sampling interval. 

The algorithm can execute simulations of considerable size in reasonable times. For example, an IBM with 2 quad-core processors (1.86GHz cores) and 2.0GB of RAM completed a $10^5 s$ simulation of the network presented in Section~\ref{TDPDs} for 8000 cells in $81 s$ when $v_{0}=0.3 s^{-1}$, $v_1=0.05 s^{-1}$, $d_0=0.05 s^{-1}$, $d_1=5 \times 10^{-5} s^{-1}$, $t_{div}=3600 s$, and $t_{restore}=3300 s$.

\section{Numerical Results}

In order to evaluate the accuracy of the present algorithm, we compare  
simulation results to steady-state and time-dependent analytical  
solutions of constitutive gene expression models. In this section,  
models describing increasingly realistic biological features are  
considered and presented along with the derivations of the  
corresponding analytical solutions. We have included these details to  
emphasize the significant complexity associated with the derivation of  
even simple kinetic models. Part of our motivation for developing the  
algorithm is the anticipation that finding analytical solutions to  
models incorporating complex biochemical reaction network and cellular  
physiology will be intractable. We begin in Subsection 5.1 by  
considering time-dependent gene expression, i.e., the transcription of  
RNA and translation of RNA into protein, and benchmark this scenario  
against the corresponding time-dependent analytical distributions. In  
Subsection 5.2 we consider both time-dependent and time-independent  
gene expression using a model that incorporates the effects of gene  
duplication and cell division on gene expression dynamics in  
individual cells using the constant-number MC method. All simulations  
statistics were obtained from populations consisting of 8000 cells.

\subsection{Time-Dependent Population Distributions} \label{TDPDs}

Population-based simulation algorithms have the advantage of yielding  
time-dependent population-distributions as the output. To evaluate the  
accuracy of our approach in this respect, validation against a time-dependent
distribution is of interest. For this purpose, we simulate a  
two-stage gene expression model consisting of the following  
biochemical reactions:
\begin{eqnarray}
T    & \stackrel{v_{0}}{\longrightarrow} & T + mRNA\label{distr_eq1}\\
mRNA & \stackrel{d_{0}}{\longrightarrow} & \oslash  \label{distr_eq2}\\
mRNA & \stackrel{v_{1}}{\longrightarrow} & mRNA + P\label{distr_eq3} \\
P    & \stackrel{d_{1}}{\longrightarrow} & \oslash \label{distr_eq4}
\end{eqnarray}
where Eq.~(\ref{distr_eq1}) describes transcription at a rate $v_{0}$,  
Eq.~(\ref{distr_eq2}) the degradation of the mRNA at a rate $d_{0}$,  
Eq.~(\ref{distr_eq3}) translation at a rate $v_{1}$, and  
Eq.~(\ref{distr_eq4}) the protein degradation at a rate $d_{1}$. Here,  
all rates are given in probability per unit time and it is assumed  
that the promoter $T$ is always active and thus the model has two  
stochastic variables, the number of mRNAs and the number of proteins $P 
$.

Shahrezaei and Swain \cite{Swain3} studied the system described by  
Eqs.~(\ref{distr_eq1})-(\ref{distr_eq4}) and derived an approximative  
protein distribution as a function of time. The approximation is based  
on the assumption that the degradation of mRNA is fast compared to the  
degradation of proteins (i.e. $d_0/d_1 \gg 1$). Consequently, the  
dynamics of mRNA is at the steady-state for the most of a protein's  
lifetime. The essential steps of the derivation are as follows (see  
supplementary materials in \cite{Swain3} for the complete derivation):

The chemical master equation (CME) describing the probability of  
having $m$ mRNAs and $n$ proteins for the system in  
Eqs.~(\ref{distr_eq1}-\ref{distr_eq4}) at time $t$ is given by
\begin{eqnarray} \label{sec5.1me}
\frac{\partial P_{m,n}}{\partial t} & = & v_{0}(P_{m-1,n}-P_{m,n}) +  
v_{1}m(P_{m,n-1}-P_{m,n}) \nonumber \\
& & {} + d_{0}[(m+1)P_{m+1,n}-mP_{m,n}] \nonumber \\
& & {} + d_{1}[(n+1)P_{m,n+1}-nP_{m,n}]. \nonumber
\end{eqnarray}
If we let $u=z^{'}-1$ and $v=z-1$, the corresponding generating  
function $F(z^{'},z)$, defined in \cite{Swain3} as $\sum_{m,n} 
(z^{'})^{m}z^{n}P_{m,n}$, is given by
\begin{equation} \label{sec5.1mgf}
\frac{1}{v}\frac{\partial F}{\partial \tau} + \frac{\partial F} 
{\partial v} - \gamma\left[b(1+u)-\frac{u}{v}\right]\frac{\partial F} 
{\partial u} = a\frac{u}{v}F,
\end{equation}
where $a=v_{0}/d_{1}$, $b=v_{1}/d_{0}$, $\gamma={d_{0}/d_{1}}$, and $ 
\tau=d_{1}t$. If $r$ measures the distance along a characteristic,  
which starts at $\tau=0$ with $u=u_{0}$ and $v=v_{0}$ for some  
constants $u_{0}$ and $v_{0}$, then from Eq. \ref{sec5.1mgf} it is
found that
\begin{equation} \label{sec5.1eqn1}
\frac{du}{dr} = -\gamma\left[b(1+u)-\frac{u}{v}\right]
\end{equation}
using the method of characteristics. Consequently direct integration  
implies that $v=r$ and Eq. \ref{sec5.1eqn1} has the solution
\begin{equation} \label{sec5.1eqn2}
u(v) = e^{-\gamma bv}v^{\gamma}\left[C-b\gamma \int^{v} dv^{'}  
\frac{e^{\gamma b v^{'}}}{v^{'\gamma}} \right]
\end{equation}
for a constant $C$. By Taylor expansion of $e^{\gamma bv}$ such that  
$e^{\gamma bv}=\sum_{n}(\gamma bv)^{n}/n!$ the integral in Eq.  
\ref{sec5.1eqn2}
can be evaluated, and, if Stirling's approximation is subsequently  
applied, $u(v)$ is found for $\gamma>>1$ to obey
\begin{equation} \label{sec5.1eqn3}
u(v) \cong \left( u_{0} - \frac{bv_{0}}{1-bv_{0}} \right) e^{-\gamma  
b(v-v_{0})} \left( \frac{v}{v_{0}} \right)^{\gamma} + \frac{bv}{1-bv}
\end{equation}
or
\begin{equation} \label{sec5.1eqn4}
u(v) \cong \frac{bv}{1-bv}
\end{equation}
as $v=v_{0}e^{\tau}>v_{0}$ for $\tau > 0$. When $\gamma>>1$, $u$ tends  
rapidly to a fixed function of $v$ and the generating function  
describing the distribution of proteins can be obtained from Eq.  
\ref{sec5.1mgf}
\begin{equation} \label{sec5.1eqn5}
\frac{dF}{dv} \cong \frac{ab}{1-bv}F.
\end{equation}
Integrating Eq. \ref{sec5.1eqn5} yields the probability distribution  
for protein number as a function of time
\begin{equation} \label{sec5.1eqn6}
F(z,\tau)=\left[\frac{1-b(z-1)e^{-\tau}}{1+b-bz}\right]^{a}.
\end{equation}
By definition of a generating function, expanding $F(z)$ in $z$ yields
\begin{equation}\label{distr_eq5}
         P_{n}(\tau) = \frac{\Gamma(a + n)}{\Gamma(n + 1)\Gamma(a)}
         \left[\frac{b}{1 + b}\right]^{n}
         \left[\frac{1 + be^{-\tau}}{1+b}\right]^{a}
         \times\vspace{0.5mm}_{2}F_{1}\left[-n,-a,1-a-n;\frac{1+b} 
{e^{\tau} + b}\right],
\end{equation}
where $_{2}F_{1}$ and $\Gamma$ are the hypergeometric and the gamma  
function, respectively. The initial number of proteins $n$ is set to  
zero. In this case, the mean, variance, and protein noise of the  
process are described respectively by
\begin{equation}\label{distr_eq6}
\mu_{P}(\tau)=ab(1-e^{-\tau}),
\end{equation}
\begin{equation}\label{distr_eq7}
\sigma^{2}_{P}(\tau)=\mu_{P}(1+b+b e^{-\tau}),
\end{equation}
\begin{equation}\label{distr_eq8}
\eta_P(\tau)={\sigma_{P}}/{\mu_{P}}=\left[ \frac{1+b+be^{-\tau}}{ab(1- 
e^{-\tau})} \right]^{1/2}.
\end{equation}

\begin{figure}[h]
\begin{center}
\includegraphics*[width=14cm,height=10cm]{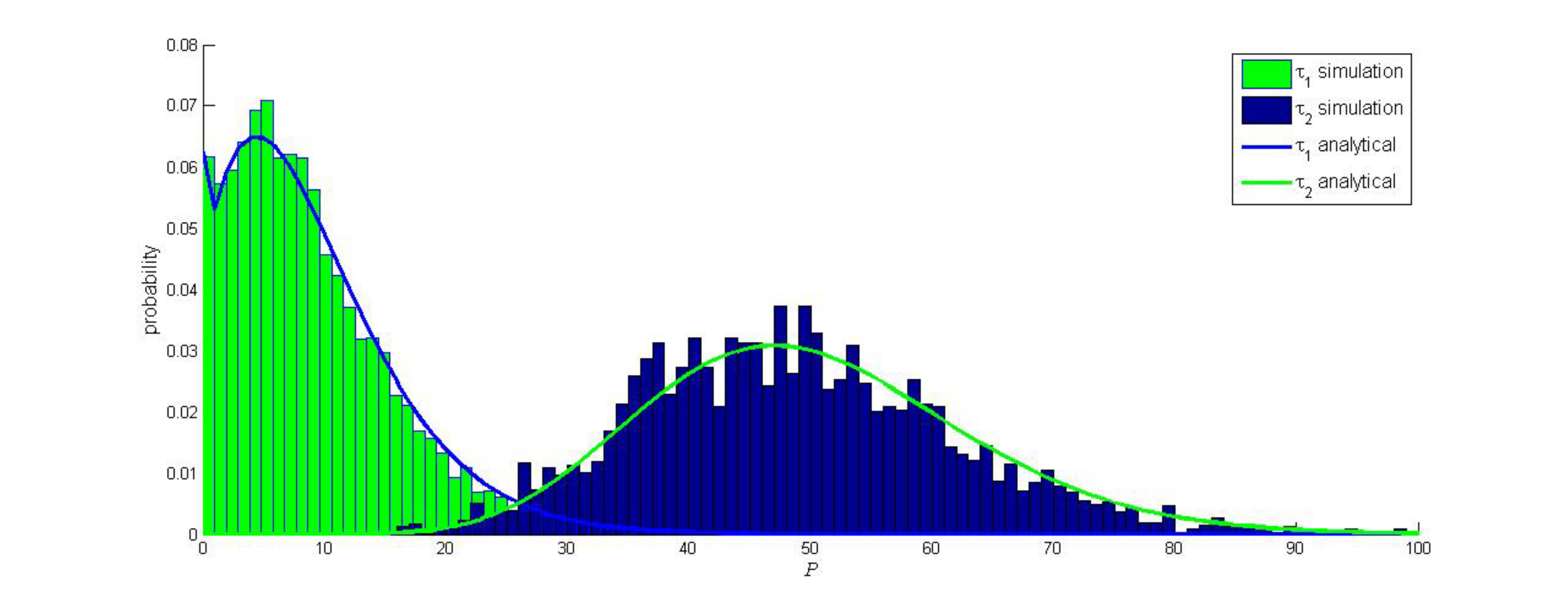}
\end{center}
\caption{\protect
     Simulation results and time-dependent analytical solutions of a  
two-stage model of gene expression \cite{Swain3}. The distribution of  
protein numbers for a population of cells at two different  
dimensionless times, $\tau=0.2$ and $\tau=10$, is shown.
} \label{swain_pop_hist}
\end{figure}

\begin{figure}[h]
\begin{center}
\includegraphics*[width=14cm,height=10cm]{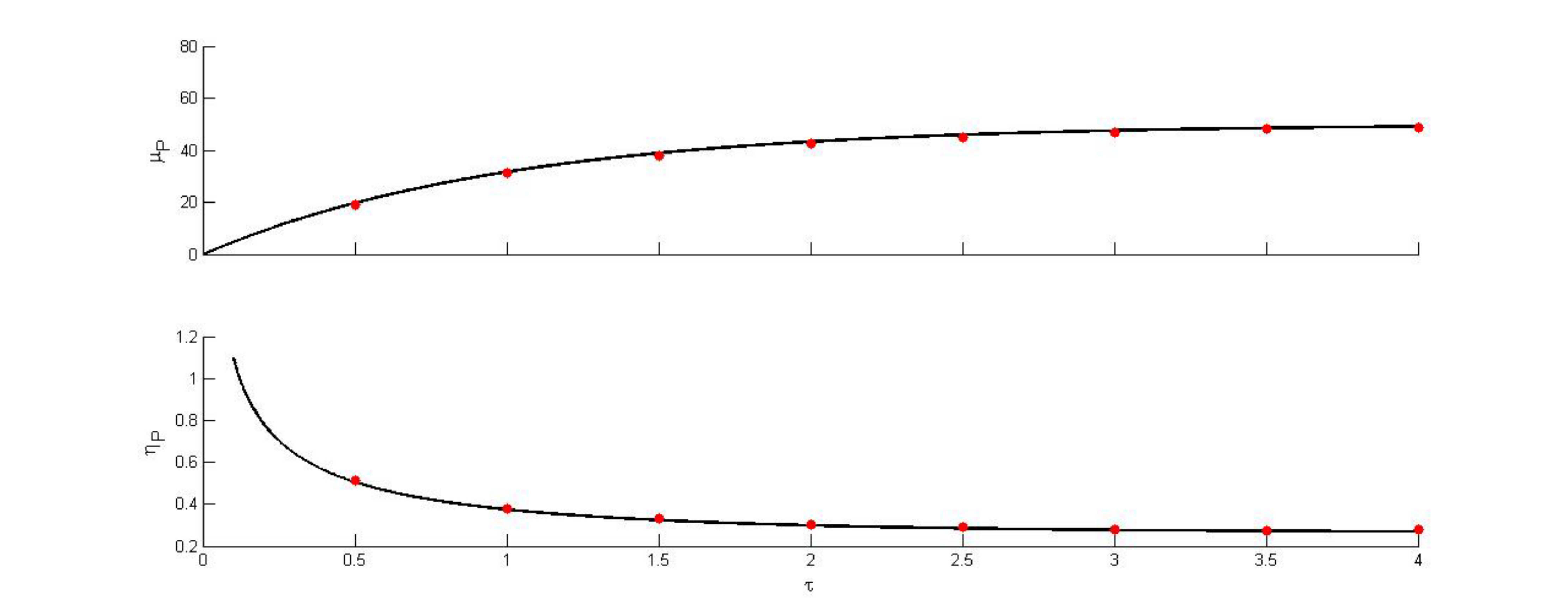}
\end{center}
\caption{\protect
     Simulation results and time-dependent analytical solutions of a  
two-stage model of gene expression \cite{Swain3}. Mean protein $\mu_{P} 
$ (top) and noise $\eta_{P}$ (bottom) are plotted as a function of  
dimensionless time $\tau$. Red dots indicate simulation results and  
black curves analytical solutions \cite{Swain3}.
} \label{swain_pop_stats}
\end{figure}

To benchmark the ability of the algorithm to accurately generate time-dependent
population distributions, we simulated  
Eqs.~(\ref{distr_eq1})-(\ref{distr_eq4}) under conditions where the  
assumptions of Eq.~(\ref{distr_eq5}) are satisfied, and compared the  
resulting distributions with corresponding time-dependent analytical  
distributions. Fig. \ref{swain_pop_hist} shows the simulated and  
analytical distributions at two different values of dimensionless time  
$\tau$. The population statistics, specifically $\mu_{P}$ and $\eta_{P} 
$, as a function of $\tau$ are shown in Fig. \ref{swain_pop_stats}. In  
both cases, the simulated protein distributions and statistics are in  
excellent agreement with the analytical results (Eqs.  
(\ref{distr_eq5})-(\ref{distr_eq8})).

\subsection{Gene Duplication, Cell Division, and Time-Dependent  
Validation}

To explore the accuracy of the algorithm when simulating models  
incorporating cell growth, division, and DNA replication, we  
implemented the simplified reaction network presented in Swain {\it et  
al.}~\cite{Swain}. The reduced reaction network was obtained from a  
model of gene expression consisting of $8$ molecular species and $11$  
chemical reactions. For this simplified network, it is possible to  
derive time-dependent analytical results for the mean protein number  
and coefficient of variation in protein number. Importantly, by making  
the appropriate approximations, the effects of gene replication and  
cell division can be included in the analytical solutions. The reduced  
model have two components - one described by the reactions in  
Eqs.~(\ref{distr_eq1})-(\ref{distr_eq4}) (note that the reaction rates
$v_{1}$ and $d_{0}$ can be directly related to $v^{'}_{1}$ and $d^{'}_{0}$
in the original model \cite{Swain}), and another describing pre-transcription
kinetics. This component captures the reversible binding  
of RNAP to the promoter (rate constants $b_0$ and $f_0$), and the  
formation of an  open promoter complex (rate constant $k_0$). These  
steps are described by the reactions
\begin{eqnarray}
        D                 &        \stackrel[b_{0}]{f_{0}}{\rightleftharpoons} & C                                 
\label{swain2002react1}\\
        C                 &        \stackrel{k_{0}}{\longrightarrow} & D + T                     
\label{swain2002react2}
\end{eqnarray}
where $D$, $C$ and $T$ represent the promoter with polymerase unbound, the promoter  
with polymerase bound and the open promoter complex, respectively. Since  
the total number $n$ of DNA molecules is conserved before and after  
replication, $D$ and $C$ can be constrained by
\begin{equation}\label{DNAconstraint}
        n_0 + n_1 = n,
\end{equation}
where $n_0$ and $n_1$ are the number of promoter copies in state $D$  
and $C$ respectively.

To derive an analytical solution, the authors invoked the assumption  
that the distributions of $C, T$, and $mRNA$ can be approximated by  
their steady state distributions. While this assumption thus ignores  
the transient dynamics of these species, it is expected to introduce a  
minimal error since the protein degradation rate $d_{1}$ is much  
smaller compared to the other reaction rates. As a consequence, the  
mean and coefficient of variation protein $P$ are time-dependent while  
the moments of the distributions of the other species are constant.  
Even with this approximation, the derivation of the analytical  
solutions for the mean and coefficient of variation is rather arduous.  
In the following, we highlight the only the main points (the complete  
derivation can be found in the supplementary material of Swain et al.  
\cite{Swain}). It consists of three separate stages - the derivation  
of time-dependent expression for the population mean and noise, the  
incorporation of gene replication and the addition of cell division.

The first stage is analogous to the derivation of time-dependent  
moments in Section~\ref{TDPDs}, that is, cell cycle effects are  
neglected and the probability distributions for the species $C, T,  
mRNA,$ and $P$ is described using the CME. In this case, the variables  
$n_{1} , n_{2}, n_{3}$, and $n_{4}$ are used to describe the numbers  
of $C, T, mRNA,$ and $P$, respectively, and  
$p(n_{1},n_{2},n_{3},n_{4},t)$ denotes the probability density  
function of the time-dependent state. The CME can be correspondingly  
be written in the form
\begin{eqnarray} \label{sec5.2me}
        \frac{\partial p(n_{1},n_{2},n_{3},n_{4},t)}{\partial t} & = & f_{0} 
[(n-n_{1}+1)p(n_{1}-1,n_{2},n_{3},n_{4},t)\nonumber \\
        &  &-(n-n_{1})p(n_{1},n_{2},n_{3},n_{4},t)] + \cdots,
\end{eqnarray}
where dots denote similar terms, one for each rate constant. The CME  
is then used to derive an expression for the time-dependent  
probability-generating function. The probability-generating function  
is defined by
\begin{equation}\label{sec5.2probgenfunc}
        F(z_{1},z_{2},z_{3},z_{4},t) = \sum_{n_{1},n_{2},n_{3},n_{4}}  
z_ 
{1 
}^ 
{n_ 
{1}}z_{2}^{n_{2}}z_{3}^{n_{3}}z_{4}^{n_{4}}p(n_{1},n_{2},n_{3},n_{4},t).
\end{equation}
It can easily be seen that differentiating $F$ with respect to $z_{i}$  
and setting all $z_{i}$ to unity, gives $\mu_{n_{i}}$ and similarly  
the second derivative gives $\mu_{n_{i}(n_{i}-1)}$. Applying the  
transformation given by Eq. (\ref{sec5.2probgenfunc}) to the CME (Eq.  
(\ref{sec5.2me})), an expression for the probability-generating  
function can be obtained. This expression has the form of the partial  
differential equation
\begin{eqnarray} \label{sec5.2PDEPGF}
        \frac{\partial F}{\partial t} & = & f_{0}nwF - \left[f_{0}w(1+w) +  
b_{0}w - k_{0}(x-w)\right]\frac{\partial F}{\partial w} + v_{0}(y-x) 
\frac{\partial F}{\partial x}\nonumber \\
        & & +\left[v_{1}'z(1+y) - d_{0}'y\right]\frac{\partial F}{\partial y}  
- d_{1}z\frac{\partial F}{\partial z},
\end{eqnarray}
where $w = z_{1} -1$, $x = z_{2} -1$, $y = z_{3} -1$, and $z = z_{4}  
-1$. This equation, just like the CME, is practically impossible to  
solve. However, the equation can be combined with a second order  
Taylor expansion of Eq. (\ref{sec5.2probgenfunc}) which can be written  
in the form
\begin{eqnarray} \label{sec5.2taylor}
        F(w,x,y,z,t) & \simeq & 1 + wX_{1} + xX_{2} + yX_{3} + zX_{4}(t) +  
\frac{1}{2}\big[X_{11}w^2 + X_{22}x^2  \nonumber\\
        & & + X_{33}y^2 + X_{44}(t)z^2 + 2X_{12}wx + 2X_{13}wy + 2X_{23}xy  
\nonumber\\
        & & + 2X_{14}(t)wz + 2X_{24}(t) + 2X_{34}(t)yz\big],
\end{eqnarray}
where the expansion is taken around $w=0$, $x=0$, $y=0$, $z=0$ so that  
the following holds: $X_{i}=\mu_{n_{i}}$, $X_{ii}=\mu_{n_{i}^2} -  
\mu_{n_{i}}$, and $X_{ij}=\mu_{n_{i}n_{j}}, i\neq j$. Here it is  
important to note that only the processes involving protein molecules  
are time-dependent according to the previous assumptions. The Eq.  
(\ref{sec5.2taylor}) is then substituted to Eq. (\ref{sec5.2PDEPGF}),  
the coefficients are compared and solvable expressions for the  
expected values, variances, and covariances of the considered process  
are obtained. This gives equations governing the variables $X_{4}= 
\mu_{P}$ and $X_{44}=\mu_{P(P-1)}$
\begin{eqnarray}
        \frac{dX_{4}(t)}{dt} & = & v_{1}'X_{3} - d_{1}X_{4}(t),\label{sec5.2  
X4}\\
        \frac{dX_{44}(t)}{dt}& = & 2v_{1}'X_{34}(t) - 2d_{1}X_{44}(t). 
\label{sec5.2 X44}
\end{eqnarray}
Assuming that $\mu_{P}(0) = m$, Eqs. (\ref{sec5.2 X4}) and  
(\ref{sec5.2 X44}) can be solved using expressions for the other  
$X_{ij}$ variables. The expressions are rather complex and the  
interested reader should refer to \cite{Swain}. Solving Eqs. (\ref{sec5.2  
X4}) and (\ref{sec5.2 X44}) yields the following expressions for the  
protein mean and variance
\begin{eqnarray}
        \mu_{P}(t) & = & \frac{v_{1}X_{3}}{d_{1}}\left(1 - e^{-d_1 t}\right) 
+me^{-d_{1}t}, \label{sec5.2protMean}\\
        {\sigma}^2_{P}(t) & = & \left(1 - e^{-d_{1}t}\right)\left(me^{-d_{1}t} 
+\lambda\left[1+\lambda\Omega\left(1 + e^{-d_{1}t}\right)\right] 
\right), \label{sec5.2protVar}
\end{eqnarray}
where
\begin{equation}\label{lambda}
        \lambda = \frac{v_{1}'f_{0}k_{0}n}{d_{0}'d_{1}l},
\end{equation}
and
\begin{equation}\label{omega}
        \Omega = \frac{d_{1}}{d_{0}' + d_{1}}\left[\eta_{33}^{2} +  
\frac{d_{0}'}{d_{1} + v_{0}}\left(\eta_{23}^{2} + \frac{v_{0}}{d_{1}+l} 
\eta_{13}^{2}\right)\right].
\end{equation}
Note that $\Omega$ is a measure of the mRNA fluctuations, $l=f_{0} +  
b_{0} + k_{0}$, and that
$\eta_{ij}^2$ is given by
\begin{equation}\label{eta}
        \eta_{ij}^2 = \frac{\mu_{n_{i}n_{j}} -\mu_{n_{i}}\mu_{n_{j}}} 
{\mu_{n_{i}}\mu_{n_{j}}}.
\end{equation}

The effects of gene replication are incorporated in the second stage  
of the derivation. The number of proteins at the beginning of each  
cell cycle is determined by the time evolution of the system during  
the cycle of a parent cell. To assess the time evolution of protein  
molecules during the cell cycle, the probability $q_{n|m}(t)$ of  
having $n$ proteins at time $t$, given that there were $m$ proteins at  
time $t=0$ is defined and the probability-generating function $Q_m(z,t) 
$ for this distribution is constructed. By definition, the generating  
function has the form
\begin{equation}\label{genfuncQ}
        Q_m(z,t) = \sum_{n}q_{n|m}(t)z^n.
\end{equation}
The equation can be expanded around $z = 1$ which yields
\begin{equation}\label{genfuncQexpand}
        Q_m(z,t) \cong 1 + (z-1)\mu_{P} + \frac{1}{2}(z-1)^2[\mu_{P^2}- 
\mu_{P}]+\cdots
\end{equation}
This function can be determined up to the necessary level by means of  
equations $\mu_{P}(t)$ and ${\sigma}^2_{P}(t)$. Using  
Eq.~\ref{genfuncQexpand}, it is obtained that
\begin{equation}\label{genfuncQsolved}
        Q_m(z,t) = Q_0(z,t)\left[1 - e^{-d_{1}t} + ze^{-d_{1}t}\right]^m.
\end{equation}
Because the gene replication occurs at time $t = t_{d}$, two different  
forms of $Q_m(z,t)$ have to be considered: $Q_m^{(1)}(z,t)$ which is  
valid when the gene number is $n$, and $Q_m^{(2)}(z,t)$ which is valid  
when the gene number is $2n$. Thus
\begin{equation}\label{genfuncQsolved2}
        Q_m^{(i)}(z,t) = Q_0^{(i)}(z,t)\left[Y + z(1-Y)\right]^m,
\end{equation}
where $Y = 1 - e^{-d_{1}t}$. Now it is possible to proceed to the  
third stage of the derivation where cell division is included.

The third stage incorporates cell division. Cell division is in the  
model assumed to occur at fixed intervals given by the division time  
$T_d$. When $t=T_d$ it is assumed that each protein has a 50 \%  
probability of being kept in this cell (symmetric division) and the  
probability of having $n$ proteins immediately after the division is  
the binomial
\begin{equation}\label{binomial}
        {m \choose n}2^{-m}
\end{equation}
given that there are $m$ proteins just before cell division. The  
transfer probability from one cell cycle to another can be constructed  
by combining the binomial distribution with the protein distribution  
derived earlier (Eq.~\ref{sec5.2taylor}). After many divisions, the  
protein number tends to a limit cycle and expressions for the mRNA and  
protein mean and coefficient of variation can be obtained in the limit  
$d_{1}/d_{0}'\ll 1$. Through a fairly complicated set of steps, it can  
be shown \cite{Swain} that the mean mRNA number before gene  
duplication ($t < t_{d}$), and the mRNA coefficient of variation are  
given by
\begin{eqnarray}
        \mu_{mRNA} & = & \frac{f_{0}k_{0}n}{d^{'}_{0}l}         
\label{swain_theor_eq1}\\
   {\eta}^2_{mRNA} & = & \frac{1}{\mu_{mRNA}}-\frac{d^{'}_{0}v_{0} 
(d^{'}_{0}+l+v_{0})}{n(d^{'}_{0}+l) (l+v_{0})(d^{'}_{0}+v_{0})}.  
\label{swain_theor_eq2}
\end{eqnarray}
The mean protein number and coefficient of variation in protein number  
as functions of time can be derived as
\begin{eqnarray}
        \mu_{P}(t) & = &\frac{v_{1}'}{d_{1}}\mu_{mRNA}\phi_{0}(t)  
\label{swain_theor_eq3}\\
        {\eta}^2_{P}(t)& = & \frac{1}{\mu_P(t)}+\frac{1}{\mu_{mRNA}}\left[1- 
\frac{f_{0}k_{0}}{l^{2}}\right]\frac{d_{1}}{d_{0}'}\phi_{1}(t), 
\label{swain_theor_eq6}
\end{eqnarray}
where
\begin{equation}\label{swain_theor_eq4and5}
\phi_{0}(t) = \left\{
        \begin{array}{ll}
           1 - \frac{e^{-d_{1}(T-t_{d}+t)}}{2-e^{-d_{1}T}},          & \: for  
\; 0 \leq t \leq t_{d} \\
                2\left[1-\frac{e^{-d_{1}(t-t_{d})}}{2-e^{-d_{1}T}}\right], & \: for  
\; t_{d} \leq t \leq T
   \end{array}
   \right.
\end{equation}
and
\begin{equation}\label{swain_theor_eq7and8}
\phi_{1}(t) = \frac{2-e^{-d_{1}T}}{2+e^{-d_{1}T}}\times\left\{
        \begin{array}{ll}
\frac{4-e^{-2d_{1}T}-2e^{-2d_{1}t}-e^{-2d_{1}(T+t-t_{d})}}{\left(2-e^{- 
d_{1}T}-e^{-d_{1}(T+t-t_{d})}\right)^{2}}, & \: for \; 0 \leq t \leq  
t_{d} \\                \frac{4-e^{-2d_{1}T}-e^{-2d_{1}t}-2e^{-2d_{1}(t-t_{d})}} 
{2\left(2-e^{-d_{1}T}-e^{-d_{1}(t-t_{d})}\right)^{2}},& \: for \;  
t_{d} \leq t \leq T.
   \end{array}
   \right.
\end{equation}
In Eqs.~(\ref{swain_theor_eq4and5}) and (\ref{swain_theor_eq7and8}),  
$t_{d}$ and $T$ denote the gene replication time and the cell division  
time, respectively.

It is noted that Eqs.~(\ref{swain_theor_eq1}) and  
(\ref{swain_theor_eq2}) are time independent and that the value of the  
mean is twice this result after gene replication occurs (i.e. when $t  
 > t_{d}$). The time independence follows from the assumption that the  
RNA is in a quasi-steady state proportional to the gene copy number $n 
$, and that all other time dependencies are absorbed into the protein  
distribution.

\begin{figure}[h]
\begin{center}
\includegraphics*[width=14cm,height=8cm]{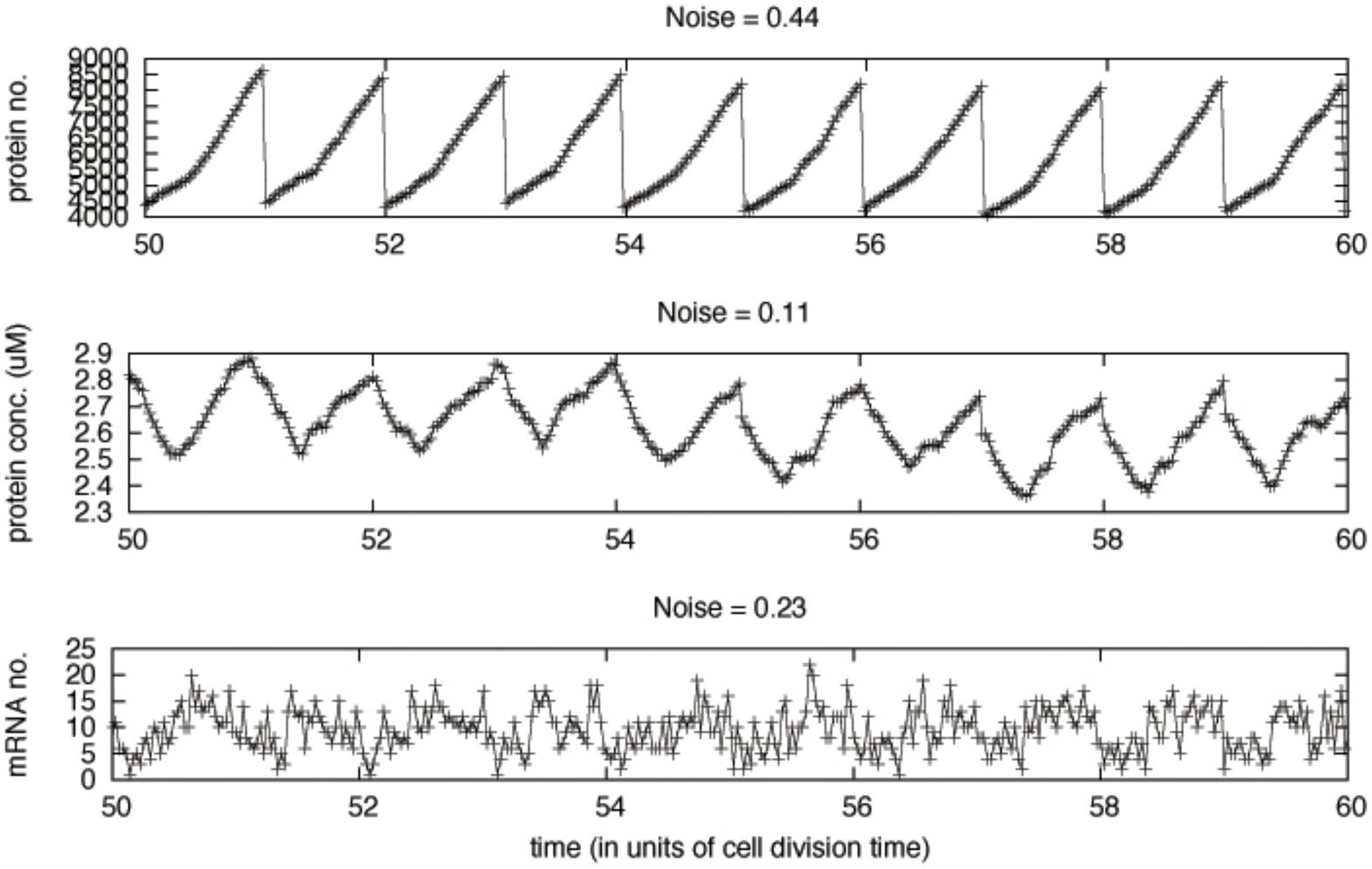}
\end{center}
\caption{\protect
     Time series of a single cell within a growing and dividing  
population. Protein number (top) and concentration (middle), and mRNA  
number (bottom), were obtained and found to be in agreement with a  
model of translation provided in \cite{Swain}. Gene duplication occurs  
every $t_{d}=0.4T$ into the cell cycle and results in an increased  
rate of protein production until the next cell division event where  
the number of genes prior to duplication is restored.
} \label{swain2_fig}
\end{figure}

\begin{figure}[h]
\begin{center}
\includegraphics*[width=14cm,height=10cm]{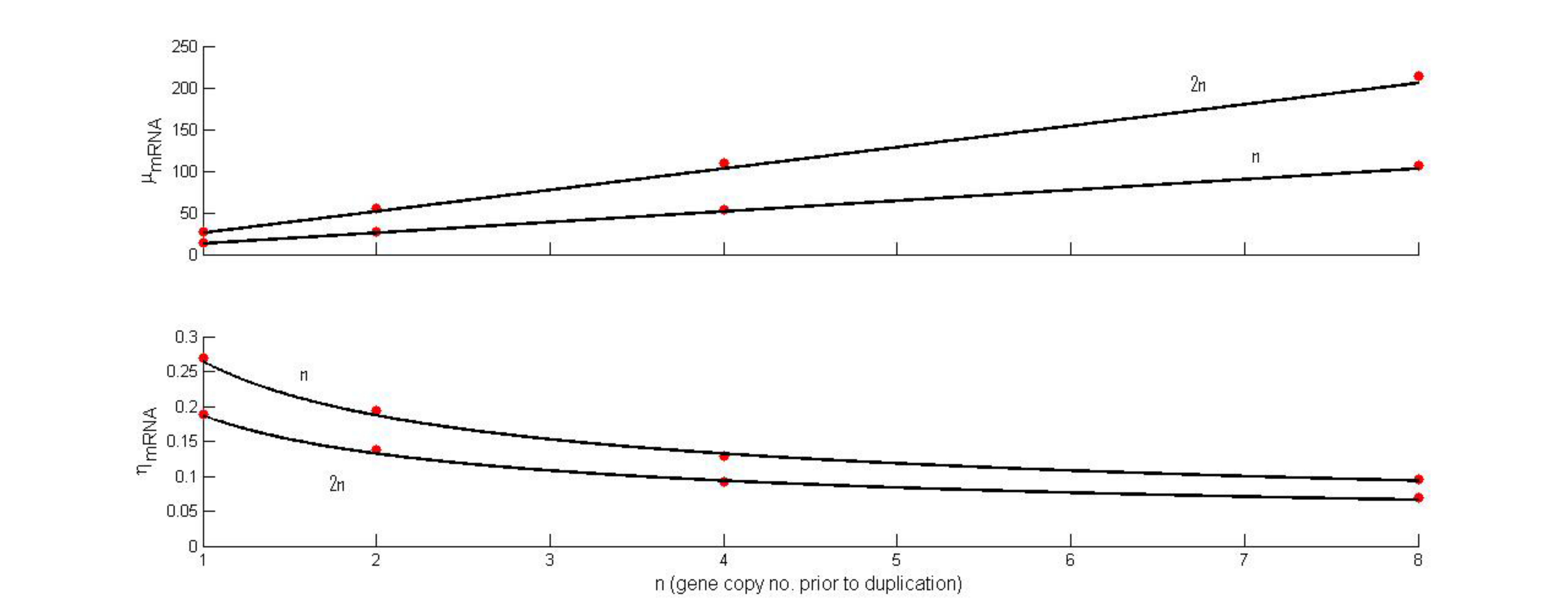}
\end{center}
\caption{\protect
     Comparison of simulation results and analytic solutions. Mean  
mRNA values are plotted as a function of gene copy number $n$ (top).  
The noise in mRNA number is also plotted as a function of $n$  
(bottom). Note that mean mRNA values increase and the noise decreases  
after gene duplication as expected. Black curves indicate analytical  
values \cite{Swain} and red dots simulation results.
} \label{swain3_fig}
\end{figure}

\begin{figure}[h]
\begin{center}
\includegraphics*[width=14cm,height=10cm]{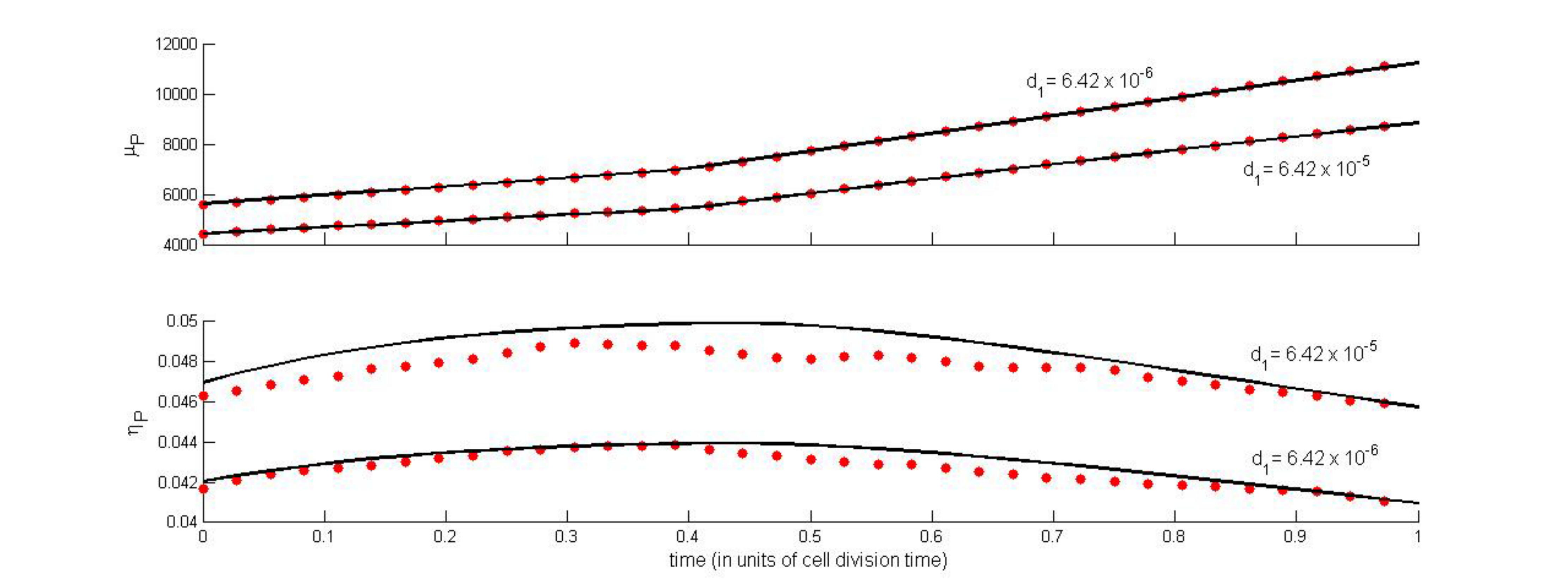}
\end{center}
\caption{\protect
     Comparison of simulation results and analytic solutions. Mean  
protein number (top) and noise (bottom) as a function of time $t$ for  
two different values of the protein degradation parameter $d_{1}$.  
Note the increase in protein production rate and decrease in noise  
levels that occurs after gene duplication at $t=0.4$. Red dots  
indicate simulation results and black curves analytical values  
\cite{Swain}.
} \label{swain4_fig}
\end{figure}

Our simulation results are compared to the corresponding steady-state  
and time-dependent analytical solutions (Figs. \ref{swain2_fig}- 
\ref{swain4_fig}). In these simulations, we use the same assumptions as  
in \cite{Swain}; the cell volume increases linearly up to time of cell  
division $T$, gene replication occurs at $t_{rep}=0.4T$ and cell  
division is symmetric with binomial partitioning of molecules.  
Simulated protein number and concentration, as well as mRNA number  
dynamics, for single cells (Fig. \ref{swain2_fig}) are comparable with  
the simulation results obtained by Swain {\it et al.}~\cite{Swain}.  
Figures \ref{swain3_fig} and \ref{swain4_fig} further compare population  
characteristics estimated from simulations to those predicted by the  
corresponding steady-state analytical solutions. Both RNA
($\mu_{mRNA}(n)$ and ${\eta}^2_{mRNA}(n)$, Fig. \ref{swain3_fig}) and  
protein ($\mu_{P}(t)$ and ${\eta}^2_{P}$, Fig.~\ref{swain4_fig})  
characteristics are in good agreement with the analytical results  
(Eqs.~(\ref{swain_theor_eq1})-(\ref{swain_theor_eq7and8})).

\section{Simulating complex population dynamics}

\subsection{Asymmetric Cell Division}

To investigate sources of external variability in eukaryotic gene expression, Volfson \textit{et al.}~\cite{Volfson} combined computational modelling with fluorescence data. As part of this study, the authors simulated the distribution of cell sizes within a population of \textit{Saccharomyces cerevisiae} (budding yeast). In these simulations, cells grew exponentially until they reached a critical volume $V_c$ where they divide. The volume at division was drawn from a normal distribution with a mean specified as a function of genealogical age and coefficient of variation 0.15. Following division, the mother cell retained 70 \% of the volume ($V_0=0.7V_c$) while daughter cells were correspondingly smaller ($V_0=0.3V_c$). The resulting distribution of cell sizes obtained from an initial population of $1000$ cells allowed to grow to $100000$ cells was found to be in agreement with experimental and analytical results \cite{Volfson}.

\begin{figure}[h]
\begin{center}
\includegraphics*[width=14cm,height=10cm]{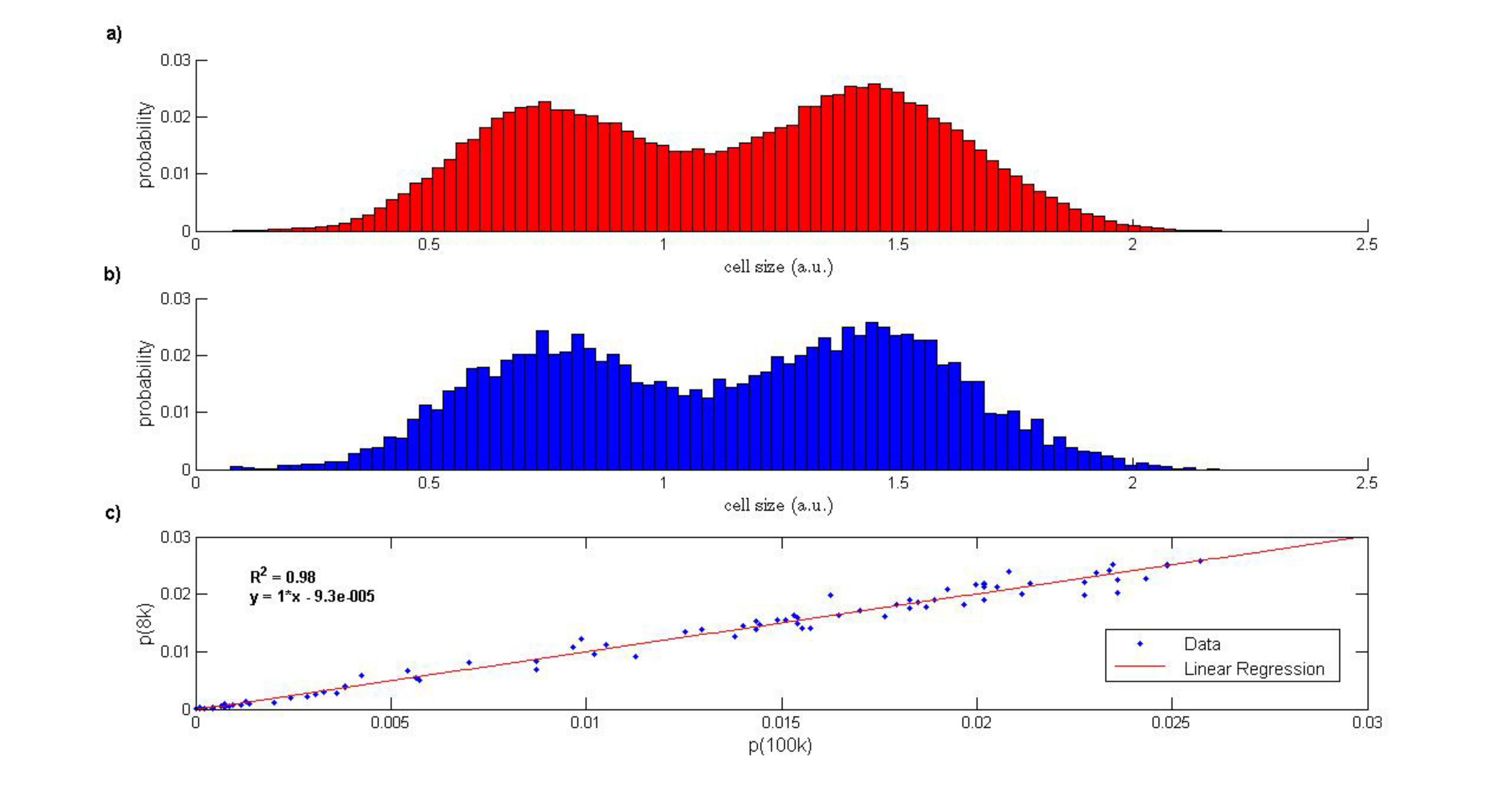}
\end{center}
\caption{\protect
     Simulation of a stochastic population dyanmics model \cite{Volfson} of a \textit{Saccharomyces cerevisiae} population undergoing stochastic (size at division) and asymmetric (partitioning of cell volume) division. (a) Steady-state distribution of cell sizes for a population of $100000$ cells. (b) Steady-state size distribution of a representative sample ($8000$ cells) obtained using the constant-number Monte Carlo method \cite{Lin,Smith} of the `true' population shown in (a). (c) Plot of the probabilities population shown in (b) against the probabilities of the population shown in (a) along with linear regression. 
} \label{cnmc_fig}
\end{figure}

The model by Volfson \textit{et al.}~\cite{Volfson} is ideally suited for benchmarking the constant-number MC method. As in Volfson \textit{et al.}~\cite{Volfson}, we first simulated the growth of a population initially consisting of $1000$ cells and obtained the steady-state size distribution once the population grew to $100000$ cells (Fig.~\ref{cnmc_fig}a). Next, we repeated the simulations using the constant-number MC method to estimate the size distribution from a representative sample ($8000$ cells) of this cell population (Fig.~\ref{cnmc_fig}b). A plot of the probabilities for the sample population against the probabilities of the `true' population shows that the difference between these variables is minimal (Fig.~\ref{cnmc_fig}c). These results compliment previous studies \cite{Lee,Lin,Mantzaris1,Mantzaris2,Smith} demonstrating the ability of the constant-number MC method to capture complex population dynamics.

\subsection{Bet-Hedging Cell Populations}

One of the most interesting potential applications of the simulation  
algorithm described in Section~\ref{sec:algorithm} is investigations  
of interactions between environmental changes, population dynamics and  
gene expression in individual cells. For example, it can be used to  
study the optimization of fitness in fluctuating environments, which is a
classic problem in evolutionary and population biology \cite{Cohen,  
Levins, Schaffer, Stearns}. Acar {\it et al.}~\cite{Oudenaarden}  
experimentally investigated how stochastic switching between  
phenotypes in changing environments affected growth rates in fast and  
slow-switching populations by using the galactose utilization network  
in \textit{Saccharomyces cerevisiae}. Specifically, a  
strain was engineered to randomly transition between two phenotypes  
($ON$ and $OFF$) characterized by high or low expression of a gene  
encoding the Ura3 enzyme necessary for uracil biosynthesis. Each  
phenotype was designed to have a growth advantage over the other in  
one of two environments. In the first environment ($E_{1}$) which  
lacks uracil, cells in the $ON$ phenotype have an advantage. In the  
second environment ($E_{2}$), cells in the $OFF$ phenotype have an  
advantage due to the presence of a drug (5-FOA) which is converted  
into a toxin by the Ura3 enzyme. In this environment, which also  
contains uracil, cells expressing Ura3 will have low viability while  
cells not expression Ura3 will grow normally.

Models of gene expression often describe the promoter $T$ as being in  
one of two states: a repressed state $T_R$ (basal level of gene  
expression) or an active state $T_A$ (upregulated level of gene  
expression) corresponding respectively to $OFF$ and $ON$ phenotypes.  
This can be described by the following biochemical reaction scheme  
\cite{Kaern}:

\begin{equation}\label{ouden_eqa}
\begin{array}{c}
k_{1}\\
T_A \rightleftharpoons T_R,\\
k_{2}
\end{array}
\end{equation}
\begin{equation}\label{ouden_eqb}
T_A \stackrel{v_{0,A}}{\longrightarrow} T_A + mRNA
\end{equation}
\begin{equation}\label{ouden_eqc}
T_R \stackrel{v_{0,R}}{\longrightarrow} T_R + mRNA
\end{equation}
\begin{equation}\label{ouden_eqd}
mRNA \stackrel{d_{0}}{\longrightarrow} \oslash
\end{equation}
\begin{equation}\label{ouden_eqe}
mRNA \stackrel{v_{1}}{\longrightarrow} mRNA  + P
\end{equation}
\begin{equation}\label{ouden_eqf}
P \stackrel{d_{1}}{\longrightarrow} \oslash
\end{equation}
where Eq.~(\ref{ouden_eqa}) describes the transitions to the $T_A$ and  
$T_R$ promoter states at rates $k_{1}$ and $k_{2}$ respectively,  
Eqs.~(\ref{ouden_eqb}) and (\ref{ouden_eqc}) the mRNA production from  
the $T_A$ (at a rate $v_{0,A}$) and $T_R$ (at a rate $v_{0,R}$) states  
respectively, Eq.~(\ref{ouden_eqe}) the protein production from mRNA  
at a rate $v_{1}$, and Eqs.~(\ref{ouden_eqd}) and (\ref{ouden_eqf})  
respectively the mRNA (at a rate $d_{0}$) and protein (at a rate  
$d_{1}$) degradation.

We first follow the approach that was used in Acar {\it et al.}~\cite{Oudenaarden}
to describe the dynamics of phenotype switching,  
where cells are in either the $ON$ or the $OFF$ state:
\begin{equation}\label{ouden_eq1}
\begin{array}{c}
k_1\\
ON \rightleftharpoons OFF\\
k_2
\end{array}
\end{equation}
In this scenario, cells randomly switch between the high and low  
expressing states at rates $k_{1}$ and $k_{2}$ (see \cite{Oudenaarden}  
for parameter values corresponding to slow and fast-switching cells).  
The growth rate (Eq.~(\ref{growth_eq})) of fit cells was set higher than  
the corresponding growth rate for unfit cells in the same environment.  
In order to avoid synchronization in the population level dynamics, we  
set $V_{div} = 2V_{0} + \xi$, where $\xi$ is a small random number
drawn from a normal distribution with zero mean and 0.2 variance.

\begin{figure}[h]
\begin{center}
\includegraphics*[width=12cm,height=10cm]{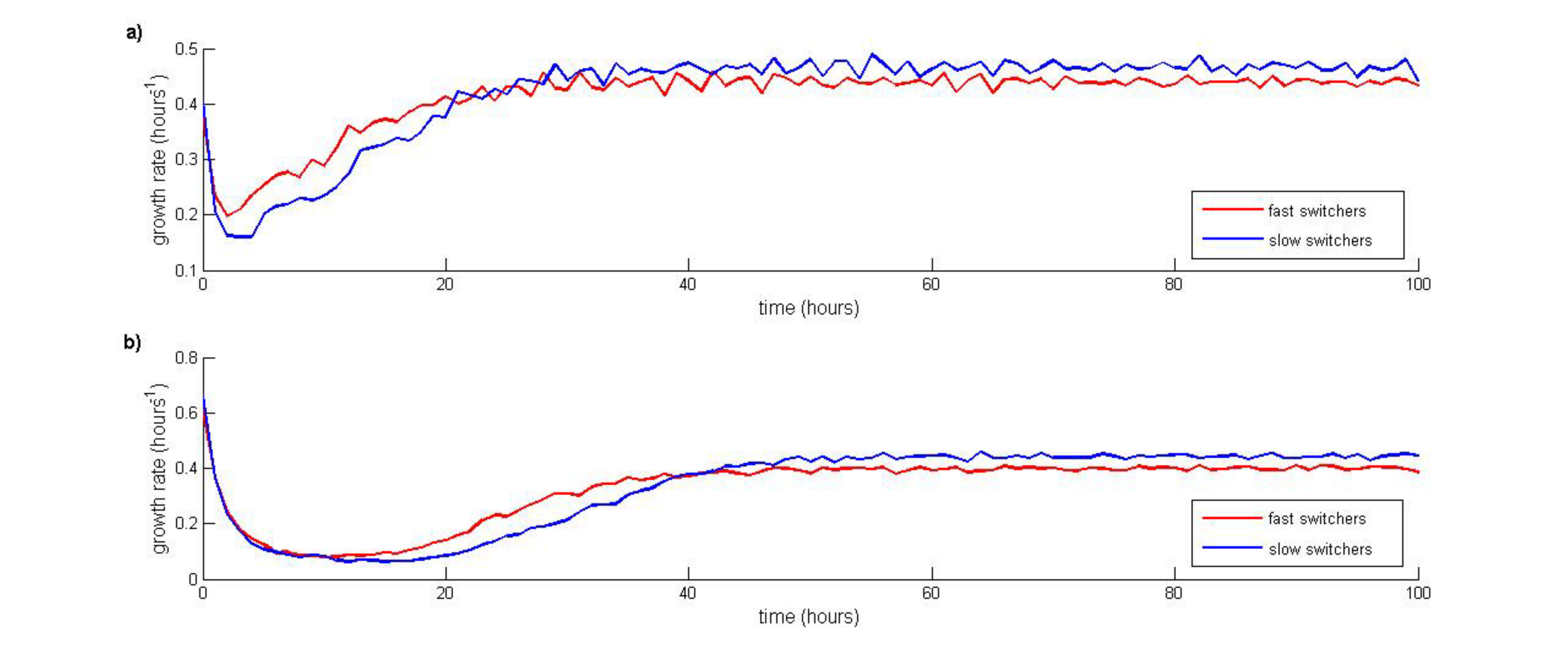}
\end{center}
\caption{\protect
     Simulations of populations of slow and fast-switching cells. (a)  
Growth rates of cells transfered from an environment containing uracil  
and 5-FOA (E2) to one containing no uracil (E1) at $t=0$. (b) Growth  
rates of cells transfered from E1 to E2 at $t=0$. Note that the  
transient before the steady-state region is shorter in (a) than in  
(b), and that fast-switching cells recover faster from the environment  
change but slow-switching cells have a higher steady-state growth, in  
agreement with experimental results found in \cite{Oudenaarden}.
} \label{ouden_fig}
\end{figure}

Figure \ref{ouden_fig} shows the growth rates obtained from  
simulations of slow and fast-switching cell populations, where cells  
were transfered from E2 to E1, and vice versa, at $t=0$. Growth rates  
show a transition period and a steady-state region. In agreement with  
experiments (see Acar \textit{et al.} \cite{Oudenaarden}), fast-switching
cells were found to recover from the effect of environment  
change faster than slow-switching cells but have a lower steady-state  
growth rate.

\begin{figure}[h]
\begin{center}
\includegraphics*[width=14cm,height=10cm]{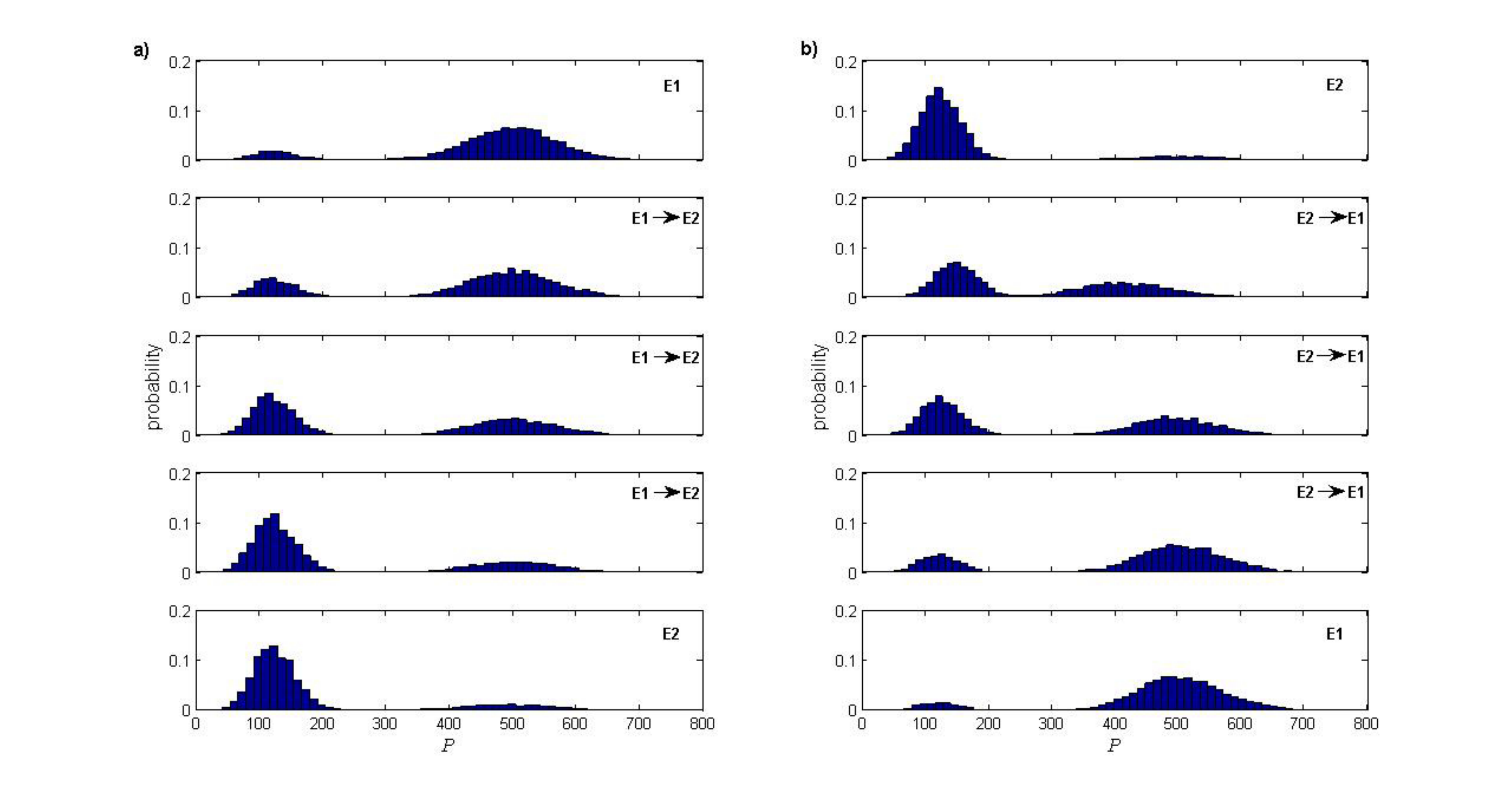}
\end{center}
\caption{\protect
     Simulations of environmental effects on phenotypic distribution.  
(a) Steady-state (top and bottom figures) and time-dependent (middle  
figures) protein distributions of cells resulting from an environment  
change from E1 to E2. (b) Steady-state (top and bottom figures) and  
time-dependent (middle figures) protein distributions of cells  
resulting from an environment change from E2 to E1. Note that when a  
sufficient amount of time has elapsed after the environmental  
transition from either E1 to E2 or vice versa, cells with either the  
OFF or ON phenotype proliferate, respectively, in agreement with  
experimental results found in \cite{Oudenaarden}. The following  
parameters were used (units $s^{-1}$): $d_0=0.005$, $v_1=0.1$,  
$d_1=0.008$, $K=200$, $n=10$. In E1 $v_{0,A}=0.2$ for fit cells and   
$v_{0,R}=0.05$ for unfit cells - vice versa in E2. Additionally $\tau_\phi$
was set to the mean doubling time (MDT) of $1.5$~hours for \textit{Saccharomyces  
cerevisiae} \cite{Brewer}.
} \label{ouden2_fig}
\end{figure}

Next we implemented the full model of gene expression described by  
Eqs.~(\ref{ouden_eqa})-(\ref{ouden_eqf}). The fitness $w_{k}$ of each  
cell $k$, which is here defined as a function of the environment and  
cellular protein concentration $[P]$, was described by a Hill function

\begin{equation} \label{fiteqn}
w_{k}(E,P) = \left\{
   \begin{array}{ll}
\frac{[P]^n}{[P]^n+K^n}, & \: if \; E = E1 \\                
\frac{K^n}{K^n+[P]^n}, & \: if \; E = E2.
   \end{array}
   \right.
\end{equation}
This equation describes partitioning of cells into fit ($w_{k}(E,P) >  
0.5$) and unfit ($w_{k}(E,P) < 0.5$) phenotypes corresponding to  
whether or not their $[P]$ in a particular environment is above or  
below a particular value given by the Hill coefficient $K$. The volume  
of each cell was described by Eq.~(\ref{growth_eq}), except here $ 
\tau_0=\tau_\phi / w$, where $\tau_\phi$ is the cell division time in  
absence of selection.  To incorporate the effect of fitness on gene  
expression, the value of transcription rate parameter $v_0$ depended  
on whether or not a cell was fit in either $E1$ or $E2$ (see Fig.~\ref{ouden2_fig}
for parameters).

The population distributions obtained for this model are shown in  
Figure \ref{ouden2_fig}. Specifically, we first obtained the steady-state
protein concentration distributions for cells in $E1$ and $E2$  
(Fig. \ref{ouden2_fig}a and \ref{ouden2_fig}b respectively). Here, the majority of  
cells either fell within a distribution centered at higher value  
characterizing the $ON$ cells, or a distribution centered at a lower  
value of $P$ characterizing the $OFF$ cells, in $E1$ or $E2$  
respectively. The rest of the cells fell within the distribution  
capturing the unfit subpopulation in both environments. These results  
were found experimentally in \cite{Oudenaarden} and are expected, as  
higher levels of the uracil enzyme are either favorable or unfavorable
with respect to the fitness of the cells depending on the environment. 
Next, the time-dependent population distributions after the transition to $E1$ from  
$E2$, and vice versa, were obtained (Fig. \ref{ouden2_fig}a and \ref{ouden2_fig}b  
respectively). Here, the dynamics of the two distinct subpopulations  
of cells in transition between the steady-states are visible. As time  
progresses after the environmental transition, less and less of the  
cells are in the unfit state ($ON$ in Fig. \ref{ouden2_fig}a and $OFF$  
in Fig. \ref{ouden2_fig}b), as the cells in the more fit state ($OFF$  
in Fig. \ref{ouden2_fig}a and $ON$ in Fig. \ref{ouden2_fig}b) grow and  
divide at a faster rate and therefore come to dominate the population  
in terms of absolute numbers.

\section{Conclusions}

We have presented a framework for the stochastic simulation of  
heterogeneous population dynamics. The accuracy of the method was  
verified by comparing simulation results of stochastic gene  
expression and population dynamics with corresponding steady-state
and time-dependent analytical solutions and experimental results.
Parallel execution of the algorithm was found to significantly
decrease run-times in comparison to simulations run on a single
processor, and did not introduce errors in numerical results.

The algorithm was also shown to be capable of simulating and capturing  
the dynamics of a cell population in a fluctuating environment, where  
phenotypic variability strongly influences gene expression dynamics.  
Agreement between this framework and the experimental and theoretical  
results obtained using a deterministic reaction-rate method in Acar  
{\it et al.}~\cite{Oudenaarden}, serves as a further benchmark for the  
proposed method. Furthermore, the algorithm's ability to capture the  
steady-state and time-independent phenotypic distributions in this  
system exemplifies the utility of this approach, as these  
distributions cannot be obtained using a deterministic framework.

Current cellular population simulation methods, including the present algorithm, treat the extracellular environment as homogeneous (e.g.~the spatial-temporal concentration profile of a nutrient required for growth is held constant). This prohibits, for example, the inclusion of competition for a limiting resource in the present implementation. However, it is possible to model feedback between cells and their environment. The simplest approach would be to assume that the environment is constant over short time intervals. The change in total population cell volume at the end of each interval could then be used to calculate how much nutrients have been consumed and the parameters describing the environment adjusted accordingly. Since the time intervals would have to be sufficiently short so that the change in concentration of the nutrient during any particular interval is negligible, the computational workload would increase substantially. The focus of future work will be on developing and benchmarking accurate and efficient augmentations that permit population simulators to handle these and other more complex scenarios. 

\section*{Acknowledgments}
This work was supported financially by the National Science and  
Engineering Research Council of Canada (NSERC), the Canadian  
Institutes of Health Research (CIHR), the Academy of Finland  
(application number 129657, Finnish Programme for Centres of  
Excellence in Research 2006-2011, and 124615), and the Tampere  
Graduate School in Information Science and Engineering (TISE).

\section*{Author Contributions}
D.C., D.F., and M.K. developed the serial, and D.C. the parallel, versions of the algorithm; D.C. performed the stochastic simulations; D.C. and J.I. benchmarked the algorithm; D.C., M.K., and J.I. wrote the paper; M.K. supervised the study.

\newpage

\newpage

\end{document}